\DocumentMetadata{}
\documentclass[acmsmall,screen]{acmart}
\usepackage{tabularx}
\usepackage{arydshln}
\usepackage{booktabs} 
\usepackage{color}
\usepackage{multirow}
\usepackage{graphicx}
\usepackage{subcaption}
\usepackage{array}
\usepackage{tcolorbox}
\usepackage{xspace}
\usepackage{colortbl}
\usepackage{mdframed}
\usepackage{enumitem}
\usepackage{appendix}
\usepackage{makecell}
\usepackage{algorithm}
\usepackage{colortbl}
\usepackage[table]{xcolor}
\usepackage{booktabs} 

\usepackage{algpseudocode}
\usepackage{listings}
\definecolor{background}{HTML}{EBFEED}
\definecolor{edge}{HTML}{B2DDA3}
\definecolor{morandiRed}{RGB}{255, 220, 220}
\definecolor{morandiGreen}{RGB}{102, 153, 136}
\definecolor{MyPurple}{RGB}{128, 0, 128}
\definecolor{airforceblue}{rgb}{0.08, 0.38, 0.74}
\newtcolorbox{mybox}{
colback=background!50, colframe=edge,
width=\columnwidth,
arc=1.2mm,
auto
outer
arc
}

\newcommand\qwenthreeeightb{\textsc{Qwen3-8B}}
\newcommand\qwenthreeforteenb{\textsc{Qwen3-14B}}
\newcommand\qwenthirtytwob{\textsc{Qwen3-32B}}

\begin{document}

\title{Empowering RepoQA-Agent based on Reinforcement Learning Driven by Monte-carlo Tree Search}
\author{Guochang Li}
\authornote{The work was done when the author was internship at Alibaba Group.}
\affiliation{%
  \institution{Zhejiang University}
  \city{Hangzhou}
  \country{China}}
\email{gcli@zju.edu.cn}

\author{Yuchen Liu}
\affiliation{%
  \institution{Alibaba Group}
  \city{Hangzhou}
  \country{China}}
\email{lyc427470@alibaba-inc.com}

\author{Zhen Qin}
\affiliation{ 
  \institution{Zhejiang University}
  \city{Hangzhou}
  \country{China}}
\email{zhenqin@zju.edu.cn}

\author{Yunkun Wang}
\affiliation{ 
  \institution{Zhejiang University}
  \city{Hangzhou}
  \country{China}}
\email{wangykun@zju.edu.cn}

\author{Jianping Zhong}
\affiliation{ 
  \institution{Zhejiang University}
  \city{Hangzhou}
  \country{China}}
\email{jpz@zju.edu.cn}

\author{Chen Zhi}
\authornote{Corresponding authors.}
\affiliation{ 
  \institution{Zhejiang University}
  \city{Hangzhou}
  \country{China}}
\email{zjuzhichen@zju.edu.cn}

\author{Binhua Li}
\affiliation{%
  \institution{Alibaba Group}
  \city{Hangzhou}
  \country{China}}
\email{binhua.lbh@alibaba-inc.com}

\author{Fei Huang}
\affiliation{%
  \institution{Alibaba Group}
  \city{Hangzhou}
  \country{China}}
\email{f.huang@alibaba-inc.com}

\author{Yongbin Li}
\authornotemark[2]
\affiliation{%
  \institution{Alibaba Group}
  \city{Hangzhou}
  \country{China}}
\email{shuide.lyb@alibaba-inc.com}

\author{Shuiguang Deng}
\affiliation{%
  \institution{Zhejiang University}
  \city{Hangzhou}
  \country{China}}
\email{dengsg@zju.edu.cn}

\begin{abstract}
Repository-level software engineering tasks require large language models (LLMs) to efficiently navigate and extract information from complex codebases through multi-turn tool interactions. Existing approaches face significant limitations: training-free, in-context learning methods struggle to guide agents effectively in tool utilization and decision-making based on environmental feedback, while training-based approaches typically rely on costly distillation from larger LLMs, introducing data compliance concerns in enterprise environments.

To address these challenges, we introduce RepoSearch-R1, a novel agentic reinforcement learning framework driven by Monte-carlo Tree Search (MCTS). This approach allows agents to generate diverse, high-quality reasoning trajectories via self-training without requiring model distillation or external supervision. Based on RepoSearch-R1, we construct a RepoQA-Agent specifically designed for repository question-answering tasks. 

Comprehensive evaluation on repository question-answering tasks demonstrates that RepoSearch-R1 achieves substantial improvements of answer completeness: 16.0\% enhancement over no-retrieval methods, 19.5\% improvement over iterative retrieval methods, and 33\% increase in training efficiency compared to general agentic reinforcement learning approaches. Our cold-start training methodology eliminates data compliance concerns while maintaining robust exploration diversity and answer completeness across repository-level reasoning tasks.

\end{abstract}


\maketitle

\section{Introduction}
\label{sec:intro}

The emergence of repository-level benchmarks such as SWE-bench~\cite{jimenez2023swe}, SWE-Lancer~\cite{miserendino2025swe}, and Repo-bench~\cite{liu2023repobench} has driven significant research into applying large language models (LLMs) to software engineering tasks. Despite these advances, LLMs face substantial challenges when addressing complex repository-level tasks, including comprehending large-scale code repositories, performing cross-file reasoning within intricate codebases~\cite{gao2025trae}, and accurately identifying code fragments relevant to specific queries.

Current approaches to repository-level tasks primarily involve designing agent workflows and tools that equip LLMs with code repository exploration capabilities. Representative frameworks include OpenHands~\cite{wang2024openhands} and SWE-Agent~\cite{yang2024swe}, which design custom agent-computer interfaces (ACI) allowing LLMs to interact with repository environments through structured actions, and Marscode Agent~\cite{liu2024marscode} for code repair tasks. These methods enable LLMs to execute multi-turn tool calls while adapting exploration strategies based on environmental feedback to achieve improved performance.

Current approaches to enhancing repository-level agents fall into two main categories. The first category focuses on training-based enhancement, where recent studies including SWE-Smith~\cite{yang2025swe}, SWE-fixer~\cite{xie2025swe}, Lingma-SWE-GPT~\cite{ma2024lingma}, and SWE-Gym~\cite{pan2024training} employ trajectory data distilled from larger LLMs (e.g., Claude~\cite{anthropic2025} and GPT~\cite{gpt412025}) for supervised fine-tuning (SFT). Reinforcement learning approaches such as SWE-RL~\cite{wei2025swe}, ReFT~\cite{luong2024reft}, and ReTool~\cite{feng2025retool} also demonstrate policy optimization through sampled experiences. However, these training-based methods all require external distillation datasets for initialization, raising significant data compliance concerns in enterprise environments.

The second category leverages inference-time sampling to generate better trajectories without requiring distilled data. Inspired by test-time scaling laws~\cite{snell2024scaling}, where increased computational effort during inference substantially enhances model output quality, recent work has applied sampling techniques to software engineering tasks. Examples include Trae Agent~\cite{gao2025trae} and SWE-Search~\cite{antoniades2024swe}, which improve repository-level program repair through extensive inference-time sampling. Techniques such as beam search~\cite{freitag2017beam} and Monte Carlo Tree Search (MCTS)~\cite{silver2016mastering} construct reasoning trees that guide models beyond default trajectories while maintaining exploration diversity. While effective, these inference-time methods require substantial computational resources for extensive sampling at each query. This dichotomy motivates our approach: integrating inference-time sampling techniques into the training process to address the data distillation dependency of training-based methods while avoiding the computational overhead of pure inference-time approaches.

Therebefore, we propose RepoSearch-R1, a novel cold-start reinforcement learning framework that integrates MCTS into the Group Relative Policy Optimization (GRPO)~\cite{shao2024deepseekmath} pipeline. Our approach generates diverse, high-quality trajectories through self-training while assigning meaningful rewards to intermediate reasoning steps. Key innovations include: (1) an exploration-decay Upper Confidence Bound for Trees (UCT) mechanism that dynamically balances exploration and exploitation, (2) a self-critique guided child-node generation process that enhances reasoning diversity and correctness, and (3) a dual-reward architecture combining LLM-based answer quality assessment with intermediate process rewards. Experimental validation on repository question-answering tasks demonstrates that RepoSearch-R1 significantly improves both QA performance and RL training efficiency, establishing its effectiveness for repository-level reasoning and navigation.

To sum up, the main contributions of this work are as follows:

\begin{itemize}[topsep=0pt]
  \item We propose RepoSearch-R1, a fully cold-start agentic RL framework that, for the first time, integrates MCTS into on-policy GRPO reinforcement learning for repository question answer tasks, generating diverse and high-quality trajectories to improve both RL performance and training efficiency.
  \item We construct RepoQA-Agent, equipped with specialized tools for reasoning and searching in code repositories, specifically designed to address repository-level question-answering tasks.
  \item We apply RepoSearch-R1 to RepoQA-Agent for agentic RL training, achieving a 19.4\% improvement over IRCoT methods, a 16.0\% improvement over naive generation, a 6.4\% improvement over RAG methods, and a 33\% increase in training efficiency compared to general agentic RL approaches. 
\end{itemize}

Training source code of this work has been publicly released on \cite{repo} to allow researchers to further extend it to other software engineering tasks. Our results demonstrate that complex reasoning abilities can be cultivated through self-training reinforcement learning without relying on distilled data from larger models. This opens a new pathway for training autonomous agents in complex, structured environments where traditional supervised learning faces challenges due to scarce data.

\section{RepoQA-Agent: Design of Repository Question Answer Agent}
\label{sec:repo_search}

Since question answering serves as an optimal evaluation paradigm for assessing retrieval effectiveness, we adopt code repository question-answering tasks as the primary validation framework. To address the QA task, we construct a RepoQA-Agent framework that equips LLMs with file and folder review capabilities, file and keyword retrieval functions, enabling them to reason about proper tool calls and search for the most relevant code to answer questions. 

\subsection{Repository Question Answer Dataset preparation}

Recent advancements in agent reinforcement learning, exemplified by studies such as Search-R1~\cite{jin2025search}, R1-searcher~\cite{song2025r1}, and ReSearch~\cite{chen2025learning}, have primarily focused on natural language scenarios. These studies train and evaluate models using multi-hop natural language question answering datasets like MuSiQue~\cite{trivedi2022musique}, HotpotQA~\cite{yang2018hotpotqa}, and 2WikiMultiHopQA~\cite{ho2020constructing}. Similarly, in the context of software code repositories, a multi-hop repository question answering dataset is essential. Multi-hop refers to the requirement to retrieve information from multiple sources within the code repository to thoroughly answer queries, as the problem descriptions alone are insufficient.

Much of the prior work in code question-answering has focused on code QA communities, utilizing datasets like CodeSearchNet~\cite{husain2019codesearchnet}, CodeQA~\cite{liu2021codeqa}, and ProCQA~\cite{li2024procqa}. In these datasets, respondents are required to understand code snippets included within the questions themselves. These snippets do not necessarily originate from a coherent code repository, and answering typically does not require repository-level exploration. Our work, however, aims to enhance the ability of repository search agents to utilize retrieval tools effectively. Consequently, we require a dataset where questions are posed about specific code repositories and answering necessitates comprehensive understanding of repository structure and content. The CoReQA~\cite{chen2025coreqa} dataset, developed from GitHub issues and comments, encompasses 176 popular repositories across four programming languages: Python, Java, Go, and TypeScript.

\begin{figure}[h]
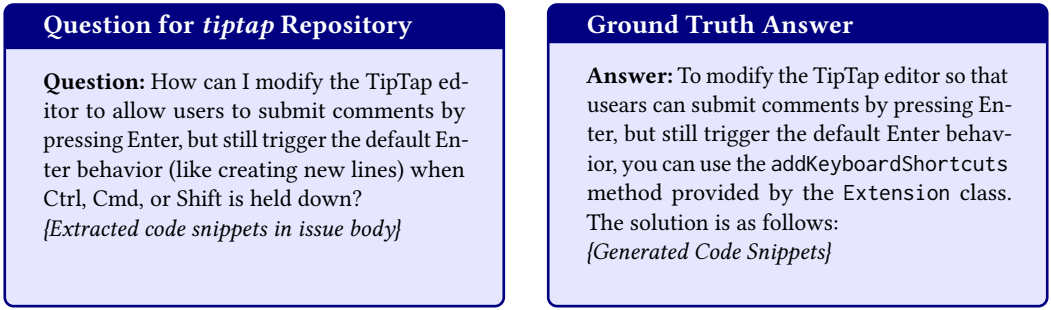

\centering
\begin{minipage}{0.48\textwidth}
\begin{tcolorbox}[
  colback=blue!10,
  colframe=blue!50!black,
  title={\textbf{Question for \textit{tiptap} Repository}},
  fonttitle=\bfseries\color{white},
  coltitle=blue!50!black,
  rounded corners,
  boxrule=1pt,
  height=4cm
]
\small
\textbf{Question:} How can I modify the TipTap editor to allow users to submit comments by pressing Enter, but still trigger the default Enter behavior (like creating new lines) when Ctrl, Cmd, or Shift is held down?\\
\textit{\{Extracted code snippets in issue body\}}
\end{tcolorbox}
\end{minipage}
\hfill
\begin{minipage}{0.48\textwidth}
\begin{tcolorbox}[
  colback=blue!10,
  colframe=blue!50!black,
  title={\textbf{Ground Truth Answer}},
  fonttitle=\bfseries\color{white},
  coltitle=blue!50!black,
  rounded corners,
  boxrule=1pt,
  height=4cm
]
\small
\textbf{Answer:} To modify the TipTap editor so that usears can submit comments by pressing Enter, but still trigger the default Enter behavior, you can use the \texttt{addKeyboardShortcuts} method provided by the \texttt{Extension} class. The solution is as follows:\\
\textit{\{Generated Code Snippets\}}
\end{tcolorbox}
\end{minipage}
\caption{QA paris example in CoReQA: repository question and ground truth answer}
\label{fig:qa_example}
\end{figure}

Figure~\ref{fig:qa_example} illustrates an example QA pair from the dataset, based on the \textit{tiptap} repository. In CoReQA, each task input consists of a user question about a repository, rewritten from a real GitHub issue. These rewritten questions may also include code snippets illustrating the user's problem. The corresponding answers are derived by rewriting comments from closed GitHub issues, ensuring that the provided information is verified and complete.

\subsection{Tool Design of RepoQA-Agent} 

We design several tools to equip the LLM with repository exploration capabilities, including viewing file directory structures, inspecting file contents, searching for files by name, and performing keyword searches. We implement these tools by encapsulating bash utilities such as \textit{cat} and \textit{grep}, creating semantically meaningful tool names: \textit{list\_files\_in\_folder}, \textit{review\_file}, \textit{search\_file\_in\_folder}, \textit{search\_symbol\_in\_file}, and \textit{search\_keyword\_in\_folder}, as detailed in Table~\ref{tab:tool_design}. This approach provides several advantages. First, it enables the model to semantically understand tool capabilities through intuitive naming conventions. In contrast to frameworks like OpenHands~\cite{wang2024openhands} and SWE-Agent~\cite{yang2024swe}, our encapsulation prevents syntax errors that occur when models use raw bash commands directly, while mitigating restrictive issues arising from excessive operational freedom. Second, since the usage patterns for these specific tools do not appear in the LLM's training data, we can authentically validate the LLM's ability to learn tool usage during reinforcement learning. Finally, we integrate access path restrictions that confine the model to repository files only, rendering all external files invisible and ensuring focused exploration.

\begin{table}[h]
  \small
  \vspace{-0.5em}
  \caption{Tools for RepoQA-Agent: These tools enable LLMs to perform repository exploration, code inspection, file navigation, and keyword/symbol searching within code repositories.}
  \label{tab:tool_design}
  \vspace{-0.3em}
  \centering
  \scriptsize
  \begin{tabular}{ccc}
      \toprule
      \textbf{Tool Name} & \textbf{Parameter} & \textbf{Tool description} \\
      \midrule

      \textbf{\texttt{review\_file}} & \texttt{file\_path, start\_lineno, end\_lineno} & \makecell{Review code in a specific file from \\ start\_lineno to end\_lineno} \\
      \cmidrule{1-3}

      \textbf{\texttt{search\_keyword\_in\_folder}} & \texttt{keyword, folder\_path} & \makecell{Search for a keyword in all files \\ within a specific folder} \\
      \cmidrule{1-3}

      \textbf{\texttt{list\_files\_in\_folder}} & \texttt{folder\_path} & \makecell{List all files and subdirectories in \\ a specific folder} \\
      \cmidrule{1-3}

      \textbf{\texttt{search\_symbol\_in\_file}} & \texttt{symbol, file\_path} & \makecell{Search for a code symbol (such as a function or \\ variable name) in a specific file} \\
      \cmidrule{1-3}

      \textbf{\texttt{search\_file\_in\_folder}} & \texttt{file\_name, folder\_path} & \makecell{Search for a specific file in all subdirectories \\ within a specific folder} \\
      \bottomrule
  \end{tabular}
\end{table}

\subsection{Multi-turn react-based tool call for RepoQA-Agent}
The RepoQA-Agent employs a ReAct framework~\cite{yao2023react} to explore code repositories through multi-turn interactions. Figure~\ref{fig:react_example} presents an example of a single exploration round. Each interaction round follows a structured three-stage process:

\textbf{Thought Phase:} The agent considers the given question, analyzes the current situation, and formulates a plan for the next action by reasoning about what information is needed and which tool would be most appropriate for collecting that information.

\textbf{Action Phase:} Based on the preceding reasoning, the agent selects and calls an available repository exploration tool (as defined in Table~\ref{tab:tool_design}). The tool is invoked with appropriate parameter configurations to search for relevant information.

\textbf{Observation Phase:} While the first two phases are generated within a single model response, the observation phase involves parsing the agent's tool call according to predefined rules, executing the tool accordingly, and returning the execution results. This stage also includes checking whether the agent process ending tag appears and monitoring whether the maximum allowed number of iterations has been reached. The resulting information is then fed back to guide the RepoQA-Agent's reasoning and decision-making in subsequent rounds.

\begin{figure}[h]
  \centering
  \begin{minipage}{0.48\textwidth}
  \begin{lstlisting}[
  basicstyle=\scriptsize\ttfamily,
  breaklines=true,
  breakindent=0pt,
  showstringspaces=false,
  rulecolor=\color{MyPurple},
  framerule=1pt,
  caption={Thought-Action pattern},
  label={fig:react_thought_action},
  xleftmargin=0pt
  ]
### Thought: 
I need to search for the keyword 'regplot' within the folder '/testbed/seaborn__569/seaborn' to find relevant code or documentation about the `regplot` function.

### Action:
```bash
search_keyword_in_folder -k 'regplot' -p '/testbed/seaborn__569/seaborn'
```
  \end{lstlisting}
  \end{minipage}
  \hfill
  \begin{minipage}{0.48\textwidth}
  \begin{lstlisting}[
  basicstyle=\scriptsize\ttfamily,
  breaklines=true,
  breakindent=0pt,
  showstringspaces=false,
  rulecolor=\color{MyPurple},
  framerule=1pt,
  caption={Tool execution result},
  label={fig:react_obs},
  xleftmargin=0pt
  ]
### Observation: 
Tool search_keyword_in_folder search result:
Found 2 matches for keyword 'regplot' in '/testbed/seaborn__569/seaborn/distributions.py' (2 matches)
Found 9 matches for keyword 'regplot' in '/testbed/seaborn__569/seaborn/linearmodels.py' (9 matches)
Found 13 matches for keyword 'regplot' in '/testbed/seaborn__569/seaborn/tests/test_linearmodels.py' (13 matches)
  \end{lstlisting}
  \end{minipage}
  \caption{Example of multi-turn interaction with code repository showing Thought-Action-Observation pattern}
  \label{fig:react_example}
  \end{figure}

This iterative process continues until the agent has gathered sufficient information to answer the given question or reaches the predefined maximum number of search rounds. The ReAct framework ensures that each exploration step is purposeful; by maintaining context across multiple interactions, the agent can develop a comprehensive understanding of the user question ande related code in repositories. During multi-turn exploration, we restrict the RepoQA-Agent to using only one tool per round. This constraint allows us to control the length of each tool output with a unified parameter, thereby mitigating the risk of exceeding the LLM context window.

\section{RepoSearch-R1 Framework}

\begin{figure}[t]
  \centering
  \includegraphics[width=\textwidth]{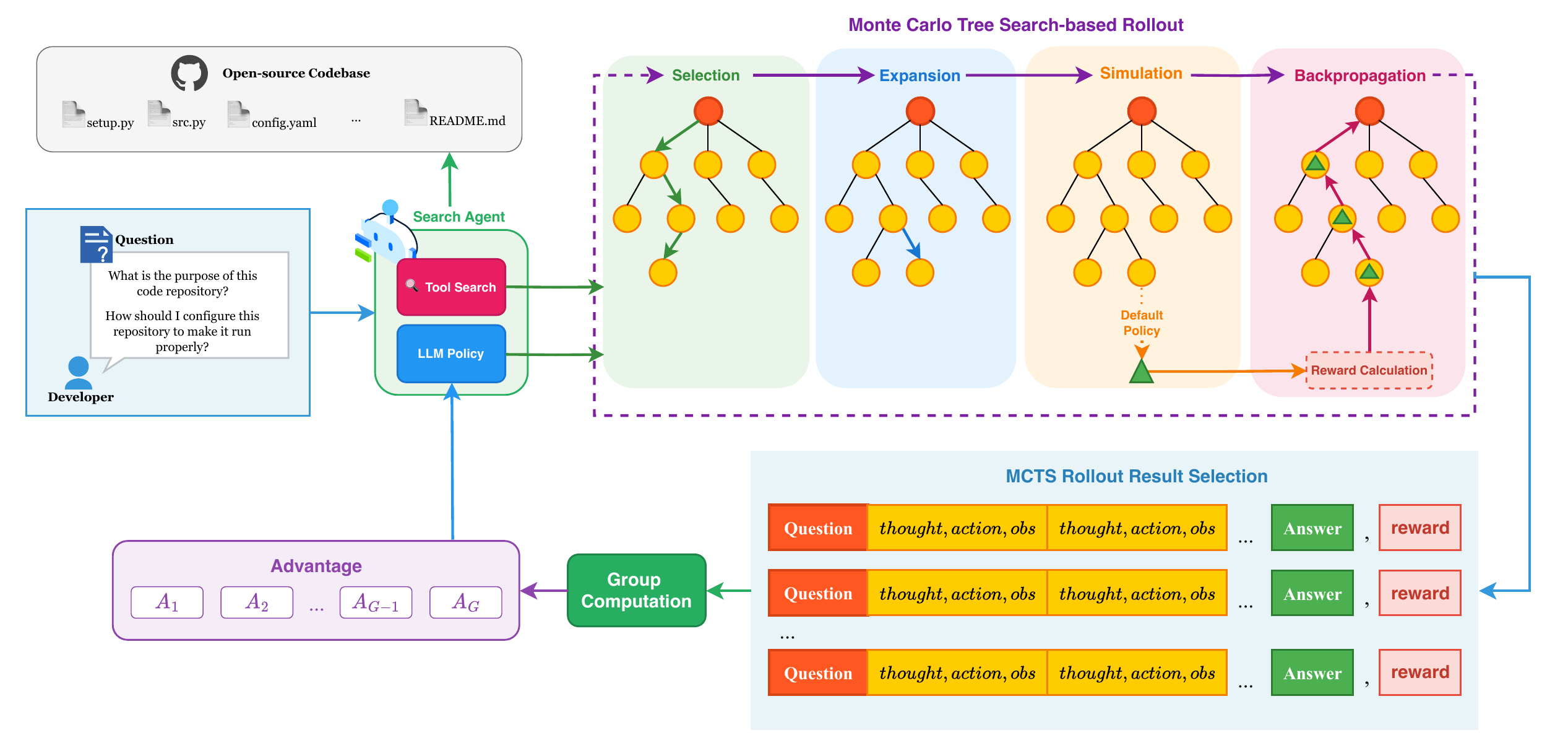}
  \caption{Overview of the RepoSearch-R1 framework showing the three-stage training pipeline: (1) MCTS-guided rollout generates diverse exploration trajectories through UCT selection, expansion, simulation, and backpropagation; (2) Trajectory selection and reward computation evaluates trajectories using reward function combining answer quality and process efficiency; (3) Advantage computation and GRPO training updates the policy using group-based normalization. The framework enables self-training agentic reinforcement learning for repository-level question answering without external supervision.}
  \label{fig:arch}
\end{figure}

The RepoSearch-R1 framework, as illustrated in Figure~\ref{fig:arch}, represents a self-training agentic reinforcement learning methodology driven by MCTS. The overall training pipeline consists of three main stages: \textbf{MCTS-guided Rollout}, \textbf{Trajectory Selection and Reward Computation}, and \textbf{ Advantage Computation and GRPO Training}. Each stage plays a crucial role in the overall learning process:

\textbf{Stage 1: MCTS-guided Rollout} The RepoQA-Agent receives a question about a code repository and performs systematic exploration using the MCTS algorithm. This process involves four key phases: Selection (choosing promising nodes using UCT), Expansion (adding new child nodes to the tree), Simulation (rollout using the current policy until terminal), and Backpropagation (updating node values with reward calculations). Multiple rollouts generate diverse exploration trajectories through self-critic and exploration-decay mechanisms.

\textbf{Stage 2: Trajectory Selection and Reward Computation} Multiple rollout trajectories are generated, each containing sequences of thought-action-observation cycles that lead to potential answers. These trajectories are evaluated using our reward function that combines LLM-based answer quality assessment with intermediate process rewards. The most promising exploration paths and their associated rewards are selected for training.

\textbf{Stage 3: Advantage Computation and GRPO Training} The selected trajectories undergo advantage computation using group-based normalization, where the relative quality of different action sequences is evaluated within each group. This information is then used in Group Relative Policy Optimization (GRPO) training to update the LLM policy, enabling the agent to make better decisions in the following training steps.

\subsection{Monte-carlo Tree Search Guided Rollout}

We next describe the specific procedures and techniques used to integrate MCTS into the reinforcement learning process. During sampling, for each question in the dataset, we maintain a Monte Carlo tree constructed through multiple rollouts. Each node in the tree represents a single interaction round and contains the thought and action generated by the model, conditioned on the chat history from the root node to its parent node. We further process the tool calls by parsing the invoked tools and appending their execution results (observations) to the chat history. Consequently, each node consists of three components: thought-action-observation.

It is important to note that the root node contains only the system prompt and the first turn user prompt including  the user's question about the repository and  question related  code snippets. In contrast, leaf nodes contain only the agent's final answer, prefixed with `\texttt{\#\#\# Answer}'.

\vspace{-0.5em}

\subsubsection{Process of Monte-carlo Tree Search}

The MCTS-guided rollout, as detailed in Algorithm~\ref{alg:mcts_grpo_unified}, constitutes a critical component of the RepoSearch-R1 framework. This process enables the RepoQA-Agent to systematically explore diverse action sequences within code repositories by leveraging a set of tools to gather information and make informed decisions. The MCTS rollout follows four sequential phases: Selection, Expansion, Simulation, and Backpropagation.

\subsubsection{Exploration-Decay UCT for MCTS selection}

The selection phase employs an Exploration-Decay Upper Confidence bound applied to Trees (UCT)~\cite{wang2024towards} formula that dynamically adjusts the exploration-exploitation balance throughout the rollout process. Unlike traditional UCT~\cite{kocsis2006bandit} that uses a fixed exploration constant, our approach implements a time-dependent exploration weight that decreases as the number of rollouts increases.

The exploration weight follows an exponential decay schedule $w(t) = w_0 \cdot (0.1)^{t/T}$, where $w_0$ is the initial exploration weight, $t$ is the current rollout index, and $T$ is the total number of rollouts. This design ensures that early rollouts prioritize exploration of diverse action sequences, while later rollouts focus more on exploiting promising paths discovered earlier. The modified UCT score is calculated as the following Equation~(\ref{eq:uct}):
\begin{equation}
  \label{eq:uct}
\text{UCT}(s, t) = \frac{q(s)}{N(s)} + w(t) \sqrt{\frac{\ln N_{parent}(s)}{N(s)}} \qquad
\end{equation}

where $q(s)$ represents the cumulative reward, $N(s)$ denotes the visit count, and $w(t)$ is the time-dependent exploration weight. Unvisited nodes are assigned infinite priority to ensure they are explored first. For the root node (which has no parent), only the exploitation term $\frac{q(s)}{N(s)}$ is used.

This exploration-decay mechanism addresses a fundamental challenge in repository exploration: early in the search process, the agent must broadly explore different tool usage patterns and repository areas, while as the search progresses, it should focus on refining the most promising exploration strategies. The exponential decay from full exploration weight to 10\% of the original value ensures a smooth transition from exploration-heavy to exploitation-focused behavior.

\begin{algorithm}[h]
  \caption{RepoSearch-R1: Self-training MCTS-guided Rollout and GRPO Training Algorithm}
  \label{alg:mcts_grpo_unified}
  \begin{algorithmic}[1]
  \Require Policy model \( \pi_{\theta} \), dataset \( \mathcal{D} \), search tools \( T \)
  \Ensure Optimized policy model \( \pi_{\theta} \)
  
  \For{each training epoch \( e = 1 \) to \( E \)}
      \For{each batch \( \mathcal{B} \) in dataset \( \mathcal{D} \)} \Comment{\textcolor{blue}{MCTS self-training Trajectory Generation}}
          \State \( Trajs \gets \emptyset \) 
          \For{each question \( q \) in batch \( \mathcal{B} \)}
              \State Initialize root node with question \( q \), Q-values, visit counts, explored nodes
              
              \For{\( i = 1 \) to \( n\_simulations \)}
                  \State \( path \gets \) Select(root) \Comment{UCT with exploration-decay}
                  \State \( node \gets path[-1] \)
                  \State Expand(\( node, \pi_{\theta} \)) \Comment{Self-critic child generation}
                  \State \( sim\_path \gets \) Simulate(\( node, T , \pi_{\theta}\)) \Comment{Tool execution}
                  \State \( complete\_path \gets path + sim\_path \)
                  \State \( reward \gets rw\_fn(complete\_path[-1]) \)
                  \State Backpropagate(\( complete\_path, reward \))
              \EndFor
          \EndFor
          \State Select trajectories from MCTS to \( Trajs \)
          \State Compute log probabilities \( \log \pi_{\theta}(a|s) \) and \( \log \pi_{ref}(a|s) \) \Comment{\textcolor{blue}{GRPO Training Update}}
          \State Group trajectories by question: \( \hat{A}_i = \frac{r_i - \mu_g}{\sigma_g + \epsilon} \)
          \State Update policy: \( \theta \leftarrow \theta + \alpha \nabla_\theta \mathbb{E}[\min(r_t(\theta) \hat{A}_t, \text{clip}(r_t(\theta), 1-\epsilon, 1+\epsilon) \hat{A}_t)] \)
          
      \EndFor
  \EndFor
  
  \State \textbf{return} optimized policy model \( \pi_{\theta} \)
  \end{algorithmic}
  \end{algorithm}
\subsubsection{Self-Critic Guided Child Generation}

The expansion phase in our MCTS implementation employs a self-critic mechanism to generate diverse and high-quality child nodes, following prior work~\cite{zhao2024marco}. Rather than generating children independently, our approach creates a first child node and then uses it as a reference to generate a second child with reflection-based prompting. The child generation process follows a structured two-step approach:

\textbf{Step 1: Standard Child Generation} The first child node is generated using the standard policy model without any additional reflection prompts. This child represents the initial response of LLM to the current state and serves as a baseline exploration path. 

\textbf{Step 2: Reflection-based Child Generation} If multiple children are required (typically 2), a second child is generated using a self-critical reflection mechanism. The reflection prompts are designed differently based on the current state of the exploration process, as shown in Figure~\ref{fig:reflection_prompts}. Agent analyzes the first child's thought-action-observation and appends a reflection prompt that encourages the LLM to reconsider its previous reasoning.

\begin{figure}[h]
\centering
\begin{tcolorbox}[
  colback=blue!5,
  colframe=blue!50!black,
  title={\textbf{Self-Critique Reflection Prompts}},
  fonttitle=\bfseries\color{white},
  coltitle=blue!50!black,
  rounded corners,
  boxrule=1pt,
  width=\textwidth
]
\small
\textbf{For Intermediate States} (when the last response was from the user):
\begin{tcolorbox}[
  colback=gray!10,
  colframe=gray!30,
  rounded corners,
  boxrule=0.5pt,
  left=2mm,
  right=2mm,
  top=2mm,
  bottom=2mm
]
\textit{``Wait! Maybe you made some mistakes! You need to rethink the last round \texttt{\#\#\# Thought} and \texttt{\#\#\# Action} and try another response.''}
\end{tcolorbox}

\vspace{2mm}
\textbf{For Terminal States} (when reaching a final answer):
\begin{tcolorbox}[
  colback=gray!10,
  colframe=gray!30,
  rounded corners,
  boxrule=0.5pt,
  left=2mm,
  right=2mm,
  top=2mm,
  bottom=2mm
]
\textit{``Wait! Maybe you made some mistakes! You need to rethink and try another answer again, remember starting with \texttt{\#\#\# Answer} Tag!''}
\end{tcolorbox}
\end{tcolorbox}
\caption{Self-Critique Reflection Prompts for Different Exploration States}
\label{fig:reflection_prompts}
\end{figure}

This self-critic approach serves multiple purposes: (1) it increases the diversity of exploration paths by encouraging the model to consider alternative reasoning strategies, (2) it helps identify potential errors in the initial reasoning process, and (3) it provides multiple perspectives on the same repository exploration problem. The reflection mechanism is particularly valuable in code repository tasks where multiple valid approaches may exist for finding relevant information, and self-correction can lead to more comprehensive and accurate solutions.

\subsubsection{Simulation and Backpropagation}

During the MCTS sampling process, the complete simulation procedure starts from the root node containing the user's question and iteratively performs node selection and expansion until a terminal node is reached. The termination condition is defined as either the agent providing an answer without performing any further retrieval or reaching the maximum allowed dialogue turns (i.e., tree depth).

After reaching a terminal node, we invoke a reward function to score the agent's generated answer against the ground truth. The resulting reward is then propagated along the exploration path, adding the reward value to each intermediate reasoning node on that path. Simultaneously, the visit count for each of these nodes is incremented by one. Note that these updated values are used in the UCT formula to guide the selection of the next node.

\subsection{KL-Free Group Relative Policy Optimization}

Group Relative Policy Optimization (GRPO)~\cite{shao2024deepseekmath} enhances the standard Proximal Policy Optimization (PPO) algorithm~\cite{schulman2017proximal} by incorporating group-based advantage estimation. Unlike PPO, GRPO does not require training a value function. Instead, it samples multiple groups of data in a single iteration and uses group-based estimations to quantify the advantage of each data point. For each question $q$ and its ground-truth answer $a$ from dataset $\mathcal{D}$, GRPO samples a group of rollout trajectories $\{o_1, o_2, \ldots, o_G\}$ from the old policy $\pi_{\theta_{\text{old}}}$ and optimizes the policy $\pi_\theta$ by maximizing the objective shown in Equation~(\ref{eq:GRPO}). Unlike the original GRPO formulation, we remove the KL divergence penalty term ($\beta \mathbb{D}_{\text{KL}}(\pi_\theta || \pi_{\text{ref}})$) to avoid constraining model diversity and enable more exploratory behavior during training.

\begin{equation}
  \label{eq:GRPO}
\begin{split}
\mathcal{J}(\theta) &= \mathbb{E}_{\mathbf{x} \sim \mathcal{D}, \{y_i, f_{i=1}^G\} \sim \pi_{\theta_{\text{old}}}(\cdot|x)} \Bigg[ \frac{1}{G} \sum_{i=1}^G \min \Bigg( \frac{\pi_\theta(y_i|x)}{\pi_{\theta_{\text{old}}}(y_i|x)} A_i, \text{clip} \left( \frac{\pi_\theta(y_i|x)}{\pi_{\theta_{\text{old}}}(y_i|x)}, 1-\epsilon, 1+\epsilon \right) A_i \Bigg)\Bigg]
\end{split}
\end{equation}

where the advantage function $A_i$ is computed using group-based normalization, for each trajectory $i$ within a group of $G$ rollouts, the advantage is calculated as the standardized deviation from the group mean reward: $A_i = (r_i - \text{mean}(\{r_j\}_{j=1}^G)) / \text{std}(\{r_j\}_{j=1}^G)$. This group-relative advantage estimation helps stabilize training by normalizing rewards within each batch, reducing variance while maintaining the relative ranking of trajectories based on their performance.

\subsection{Reward Design of RepoSearch-R1}

The reward function evaluating trajectories generated by MCTS rollout consists of three key components: (1) LLM-as-a-judge based outcome rewards for final answer quality assessment, (2) intermediate process rewards accumulated from tool execution throughout the reasoning trajectory, and (3) a reward aggregation mechanism that combines both reward types to guide policy optimization.

\subsubsection{LLM-as-a-Judge Outcome Reward}

For terminal nodes that produce final answers, we employ an LLM-as-a-judge~\cite{zheng2023judging} to assess answer quality from the completeness dimension, keeping the same evaluation method with CoReQA dataset~\cite{chen2025coreqa}. The judge evaluates whether the candidate answer addresses all key points in the question and measures the completeness of these key points against the ground truth answer using a structured prompt, as shown in Figure~\ref{fig:judge_template}, which presents the complete evaluation framework used to assess answer completeness. 

The LLM judge uses a strict 1-100 scoring scale for completeness evaluation, which is then mapped to discrete quality levels to provide stable training signals, where $s$ represents the raw llm-judge score on the 1-100 scale.:

\begin{equation}
R_{judge}(s) = \begin{cases}
0.0 & \text{if } s = 0 \text{ (Totally Wrong)} \\
0.2 & \text{if } 0 < s \leq 20 \text{ (largely incomplete, many critical points missed)} \\
0.4 & \text{if } 21 \leq s \leq 40 \text{ (significant omissions, partially complete)} \\
0.6 & \text{if } 41 \leq s \leq 60 \text{ (some omissions, covers most key points)} \\
0.8 & \text{if } 61 \leq s \leq 80 \text{ (minor omissions, mostly complete)} \\
1.0 & \text{if } 81 \leq s \leq 100 \text{ (fully comprehensive, no points missed)}
\end{cases}
\end{equation}

We discretize the continuous 1-100 scores into quality tiers rather than directly mapping to continuous 0-1.0 values to address evaluation uncertainty, particularly for borderline cases where distinguishing between similar scores (e.g., 45 vs. 55) becomes ambiguous. This discretization strategy enables the model to focus on distinguishing between meaningful quality differences while reducing noise from the inherent variability in LLM-based scoring. The judge framework is designed to evaluate based on completeness, even if the candidate's approach differs from the reference answer, which requires a larger LLM to understand text and code both in reference and candidate answer, thereby give a accurate score.

\begin{figure}[h]
\centering
\begin{tcolorbox}[
  colback=gray!5,
  colframe=gray!50!black,
  title={\textbf{LLM Judge Evaluation Template}},
  fonttitle=\bfseries\color{white},
  coltitle=gray!50!black,
  rounded corners,
  boxrule=1pt,
  width=\textwidth
]
\small
\textbf{System Prompt:} You are an impartial judge tasked with critically evaluating the quality of AI assistant responses to user questions. You will be provided with: 1) A user question (possibly including code), 2) A reference answer, 3) The AI assistant's answer.

\textbf{Evaluation Instructions:} Begin by thoroughly understanding the user question and reference answer, then rigorously assess the AI assistant's answer based on Completeness.

\textbf{Important Notes:} 
\begin{itemize}[leftmargin=1em, itemsep=0pt]
\item The reference answer may represent just one of many valid solutions
\item Evaluate based on factual correctness and effectiveness, even if the approach differs
\item For code questions, pay special attention to both explanation and implementation
\end{itemize}

\textbf{Completeness Scoring Guidelines:}
\begin{itemize}[leftmargin=1em, itemsep=0pt]
\item \textbf{1-20}: Largely incomplete, many critical points missed
\item \textbf{21-40}: Significant omissions, partially complete  
\item \textbf{41-60}: Some omissions, but covers most key points
\item \textbf{61-80}: Minor omissions, but mostly complete
\item \textbf{81-100}: Fully comprehensive, no points missed
\end{itemize}

\textbf{Response Format:} 
\begin{verbatim}
## Judge's Evaluation
### **Completeness**: [Your reasoning]
Final verdict is: [[Completeness: ?]].
\end{verbatim}
\end{tcolorbox}
\caption{LLM Judge Template for Answer Quality Assessment}
\label{fig:judge_template}
\end{figure}

\subsubsection{Intermediate Process Reward Accumulation}

To distinguish between trajectories that achieve the same final score but follow different exploration paths, we incorporate intermediate process rewards that accumulate throughout the reasoning trajectory. Each node in the MCTS tree accumulates tool rewards during the exploration process, reflecting the quality of individual tool usage decisions and encouraging efficient repository navigation patterns.

The intermediate process rewards $R_{tool}$ are computed as the cumulative sum of rewards obtained from each tool execution step along the trajectory. These rewards capture the effectiveness of the agent's exploration strategy by providing binary feedback: successful information retrieval operations receive a reward of +1.0, while incorrect or ineffective tool usage patterns are penalized with a reward of -1.0.

\subsubsection{Reward Aggregation Mechanism}

The final reward combines the outcome reward from the LLM judge with the accumulated intermediate process rewards through a weighted aggregation mechanism:

\begin{equation}
R_{final} = R_{answer} + 0.1 \times \frac{R_{tool}}{depth}
\end{equation}

where $R_{answer} = R_{judge}(s)$ represents the outcome reward, $R_{tool}$ is the accumulated intermediate process reward across the trajectory, and $depth$ is the trajectory length. The depth normalization ensures that longer trajectories are not unfairly penalized, while the 0.1 weighting factor carefully balances the contribution of process efficiency relative to final answer quality, ensuring that process rewards cannot alter the discrete quality grades determined by the LLM judge. This aggregation mechanism encourages the agent to not only achieve high-quality final answers but also to develop efficient exploration strategies throughout the reasoning process.

\subsection{Training Recipe}

\subsubsection{Curriculum Learning-Based Dataset Preprocess}

Preliminary analysis of the CoReQA dataset revealed that a substantial portion of QA pairs either lack direct relevance to the target code repository or can be resolved without repository exploration, such as cases involving simple environment configuration issues. These samples fail to satisfy our task requirements, which necessitate multi-hop reasoning through systematic repository exploration to derive comprehensive answers.

To address this challenge and following established practices in RL that require classifying data difficulty~\cite{he2025skywork}, we implement a curriculum learning~\cite{wang2021survey, shi2025efficient} approach by stratifying the dataset according to difficulty levels. We establish performance boundaries using Qwen2.5-Coder-7B-Instruct~\cite{hui2024qwen2} as the lower-capability baseline and Claude-3.7-Sonnet~\cite{claude37sonnet2025} as the upper-capability benchmark. Samples that Qwen2.5-Coder-7B-Instruct successfully resolves either in a single attempt or across eight IRCoT~\cite{trivedi2022interleaving} sampling runs are classified as trivial and excluded from training. Conversely, samples that remain unsolvable even by Claude-3.7-Sonnet using IRCoT are deemed excessively challenging and largely removed, though we retain 40 representative cases to maintain training diversity.

This curation process yields 830 high-quality QA pairs from the original 1,563 samples in CoReQA. We reserve an additional 160 samples for validation. The curated dataset is partitioned with 80\% allocated for training and 20\% for evaluation, resulting in 500 training samples and 170 validation samples.

\subsubsection{SFT-Free Cold-Start Reinforcement Learning}

Traditional reinforcement learning methodologies for agent tasks typically depend on supervised fine-tuning (SFT) using trajectory data distilled from larger LLMs or pre-existing datasets such as ReFT~\cite{luong2024reft} and ReTool~\cite{feng2025retool} for initialization. However, such distillation approaches present significant data compliance challenges in enterprise. Leveraging test-time scaling principles, we demonstrate that smaller models can achieve competitive performance through strategic sampling techniques, with MCTS serving as an effective test-time scaling mechanism.

RepoSearch-R1 addresses these challenges through a fully cold-start reinforcement learning approach that eliminates dependence on existing trajectory data or model distillation. Our method passes SFT initialization entirely, starting directly from a base LLM and generating trajectory data through MCTS rollouts. This design ensures data compliance while enabling the LLM to develop autonomous exploration strategies, unconstrained by externally distilled trajectory patterns.

\subsubsection{High-Temperature Sampling}

To enhance exploration diversity during the MCTS rollout process, RepoSearch-R1 employs a high-temperature sampling~\cite{he2025skywork} strategy in the rollout phase. Given the group-based nature of GRPO, the response sampling procedure directly influences the quality and diversity of each group, thereby affecting the overall learning performance. High-temperature sampling increases the stochasticity of action selection during simulation, encouraging the LLM to explore diverse tool usage patterns and repository navigation strategies. By exposing the LLM to a broader exploration patterns during training, this approach enables the development of more generalizable repository navigation skills while preventing overfitting to specific search strategies. Additionally, it helps maintain relatively high cross-entropy values, preserving greater potential for continued learning.

\subsubsection{Observation Mask-based Loss Calculation}

In Agengtic RL, loss aggregation is typically performed at either the token level or the sequence level~\cite{luong2024reft}. Since observations are determined by the environment rather than generated by the LLM's reasoning process, we introduce a loss mask for retrieved tokens to enable the agent to focus on refining its internal reasoning capabilities. This mask ensures that the policy gradient objective is computed exclusively over tokens generated by the LLM, excluding any content retrieved during the optimization process. Consequently, external tokens do not influence the loss computation, thereby preventing retrieved documents from interfering with the LLM's intrinsic reasoning and generation processes.

\section{EXPERIMENTS SETUP}
\label{sec:experiments_setup}

\subsection{Benchmarks}

We evaluate RepoSearch-R1 on the CoReQA dataset~\cite{chen2025coreqa}, which is specifically designed for repository-level code understanding tasks. The QA pairs in CoReQA require multi-hop reasoning across multiple files and functions within a code repository. Following our curriculum learning-based methodology, we filtered the original dataset to retain only high-quality QA pairs that genuinely necessitate repository exploration. The resulting curated dataset comprises 500 pairs for training, 160 pairs for validation, and 170 pairs for evaluation.

\subsection{Base LLMs}

Our experiments encompass both closed-source and open-source language models to provide a comprehensive evaluation across different model scales and capabilities. For closed-source models, we evaluate Claude-3.5-Sonnet ~\cite{claude35sonnet2}, GPT-4o ~\cite{gpt412025}, and Gemini-2.5 Pro ~\cite{gemini25pro2025}, which are widely adopted in code-related tasks due to their strong performance in programming domains. The open-source models include \qwenthirtytwob{}, \qwenthreeforteenb{} and \qwenthreeeightb{} from the Qwen series, which represents the most extensively used LLM family for code-related training and evaluation tasks in current research. This selection spans different parameter scales to demonstrate the effectiveness of our approach across various computational budgets, with particular focus on the \qwenthreeeightb~\cite{yang2025qwen3} model for detailed analysis of our reinforcement learning methodology.

\subsection{Baselines}

We compare RepoSearch-R1 against several established approaches for repository-level question answering:

\textbf{Naive Generation}: Direct question answering without any repository context or retrieval mechanisms. This approach represents the baseline performance of models relying solely on their pre-trained knowledge, serving as a lower bound for repository-level understanding tasks.

\textbf{RAG~\cite{chen2025coreqa}}: Traditional Retrieval-Augmented Generation (RAG) that segments repository files into chunks and employs BM25-based semantic similarity to retrieve relevant code snippets.

\textbf{IRCoT (Iterative Retrieval Chain-of-Thought)~\cite{trivedi2022interleaving}}: An advanced baseline that integrates iterative retrieval with chain-of-thought reasoning, representing current state-of-the-art methodologies for complex repository exploration tasks.

\textbf{Search-R1~\cite{jin2025search}}: A search-based reinforcement learning approach specifically adapted for repository exploration, providing a direct comparison to our MCTS-based methodology within the reinforcement learning paradigm.

These baselines collectively provide comprehensive coverage across different methodological paradigms: direct generation, retrieval-augmented approaches, iterative reasoning frameworks, and alternative reinforcement learning strategies.

\subsection{Metrics}

Our primary evaluation metric is \textbf{Completeness ($Comp_L$)}, which measures how thoroughly an LLM's response addresses all aspects of the given question, scored using the LLM-as-a-judge framework~\cite{zheng2023judging}. This metric is particularly well-suited for repository-level tasks where comprehensive understanding across multiple files and functions is essential for accurate assessment. Given the requirements for strong code comprehension capabilities and evaluation consistency, we employ Qwen3-Coder-480B-A35B-Instruct~\cite{yang2025qwen3}, currently the largest and highest-performing model in the Qwen3 Coder series, as our LLM-as-a-judge evaluator, maintaining a temperature of 0.2 across all experiments.

\begin{table}[h]
\centering
\begin{minipage}{0.54\textwidth}
\centering
\caption{Model and Training Configuration}
\label{tab:model_training_config}
\resizebox{\textwidth}{!}{
\begin{tabular}{cccccc}
\toprule
\textbf{\makecell{Batch\\Size}} & \textbf{Epochs} & \textbf{KL Loss} & \textbf{\makecell{Rollout\\Temp.}} & \textbf{\makecell{Validation\\Temp.}} & \textbf{\makecell{Group\\Size}} \\
\midrule
8 & 2 & No & 1.0 & 0.2 & 8 \\
\bottomrule
\end{tabular}
}
\end{minipage}
\hfill
\begin{minipage}{0.45\textwidth}
\centering
\caption{MCTS Agent Configuration}
\label{tab:mcts_config}
\resizebox{\textwidth}{!}{
\begin{tabular}{cccc}
\toprule
\textbf{\makecell{Rollouts\\Number}} & \textbf{\makecell{Max Depth\\(Turns)}} & \textbf{\makecell{Max\\Children}} & \textbf{\makecell{Exploration\\Weight}} \\
\midrule
8 & 10 & 2 & 2.0 \\
\bottomrule
\end{tabular}
}
\end{minipage}
\end{table}

\subsection{Implementation Details}

We implement RepoSearch-R1 using the veRL framework~\cite{sheng2025hybridflow} with the GRPO algorithm. Tables~\ref{tab:model_training_config} and~\ref{tab:mcts_config} summarize the key hyperparameters and configuration settings used in our experiments. For the MCTS algorithm, multiple sampling rounds are typically required to explore more effective retrieval strategies. However, to ensure fair comparison with Search-R1 in this study, we constrain MCTS to eight rollouts, generating at most eight agent execution trajectories per question.

\section{EXPERIMENTS RESULTS}
\label{sec:experiments_results}

To comprehensively evaluate RepoSearch-R1's effectiveness, we address the following research questions:

\textbf{RQ1: How effective is RepoSearch-R1 in enhancing repository understanding and question answering capabilities?} We evaluate whether RepoSearch-R1 can effectively and efficiently improve an agent's repository comprehension and performance of the repository QA taskthrough reinforcement learning. 

\textbf{RQ2: Can LLMs benefit from multi-turn tool reasoning for repository question answering?} Given that larger LLMs such as Claude and Gemini demonstrate strong tool usage capabilities, we examine their performance when employing multi-turn, tool-augmented reasoning. This evaluation allows us to assess whether our designed repository exploration tools provide meaningful benefits for repository-level question answering tasks.

\textbf{RQ3: Why does MCTS-based exploration outperform other reinforcement learning strategies?} We analyze the fundamental differences between MCTS-based exploration and general reinforcement learning methods by examining entropy dynamics during training and the diversity of sampled trajectories. This investigation aims to understand the underlying mechanisms that contribute to MCTS's superior learning performance and efficiency.

\subsection{RQ1: Effectiveness and Efficiency of RepoSearch-R1 Method}

To address RQ1, we evaluate whether RepoSearch-R1 can effectively enhance agents' ability to understand repositories and answer questions accurately, focusing on performance improvements across different repository reasoning scenarios.

Table~\ref{tab:validation_comparison} presents the significant improvements achieved by RepoSearch-R1 in enhancing repository understanding. On the challenging Qwen3-8B model, RepoSearch-R1 achieves a completeness score of 0.6306, representing a 16.0\% improvement over naive generation, a 6.4\% improvement over RAG, and a 19.4\% improvement over IRCoT. The training progression shows that RepoSearch-R1 reaches peak performance at step 80 with 0.631, while Search-R1 requires 120 steps to achieve its best performance. These results demonstrate that RepoSearch-R1 successfully strengthens the agent ability to comprehend complex repository structures and retrieve relevant information.

\begin{table}[h]
  \centering
  \caption{Performance comparison of Search-R1 and RepoSearch-R1 across training steps and max evaluation on Qwen3-8B. Step 0 represents the initial IRCoT baseline performance before reinforcement learning training.}
  \label{tab:validation_comparison}
  \vspace{-3mm}
  \resizebox{\textwidth}{!}{
  \begin{tabular}{c c c c c c c c}
  \toprule
  \multirow{2}{*}{\textbf{Method}} & \multicolumn{5}{c}{\textbf{Training Steps Performance(vs. IRCoT)}} & \multicolumn{2}{c}{\textbf{Max Evaluation vs. Baselines}} \\
  \cmidrule(lr){2-6} \cmidrule(lr){7-8}
  & \textbf{Step 0} & \textbf{Step 40} & \textbf{Step 80} & \textbf{Step 120} & \textbf{Step 124} & \textbf{vs. Naive Gen.} & \textbf{vs. RAG} \\
  \midrule
  Search-R1 & 0.528 & 0.535 (+1.3\%) & 0.591 (+11.9\%) & 0.622 (+17.8\%) & 0.578 (+9.5\%) & +14.3\% & +5.0\% \\
  \makecell{RepoSearch-R1 \\ \textit{rollout@8}} & 0.528 & 0.567 (+7.4\%) & \textbf{0.631 (+19.5\%)} & 0.578 (+9.5\%) & 0.621 (+17.6\%) & \textbf{+16.0\%} & \textbf{+6.4\%} \\
  \bottomrule
  \end{tabular}
  }
\end{table}

Compared to other agentic reinforcement learning methods such as Search-R1, under the same rollout budget (maintaining identical sample sizes to Search-R1), RepoSearch-R1 achieves an additional 1.3\% improvement over the Search-R1 baseline (0.6224), confirming that our MCTS-based approach provides superior guidance for repository exploration and reasoning without oversampling during the rollout stage.

Furthermore, Table~\ref{tab:validation_comparison} shows the progression of QA performance on the validation set as training steps increase. Although the final performance improvement of RepoSearch-R1 over Search-R1 is modest, RepoSearch-R1 demonstrates a clear advantage in training efficiency: it reaches peak performance at 80 steps, whereas Search-R1 requires 120 steps. From a training efficiency perspective relative to peak performance, RepoSearch-R1 improves efficiency by 33\% compared to Search-R1. Notably, RepoSearch-R1 surpasses Search-R1's best performance before the 80-step.
\begin{figure}[h]
  \centering
  \begin{tcolorbox}[
    colback=gray!5,
    colframe=gray!50!black,
    title={\textbf{Answer to RQ1:}},
    fonttitle=\bfseries\color{white},
    coltitle=gray!50!black,
    rounded corners,
    boxrule=1pt,
    width=\textwidth
  ]
  \small
  RepoSearch-R1 demonstrates substantial effectiveness and efficiency improvements in repository understanding and question answering tasks. Experimental results on the Qwen3-8B model show significant performance gains: 16.0\% improvement over naive generation, 6.4\% over RAG, and 19.4\% over IRCoT methods. Beyond performance gains, RepoSearch-R1 exhibits superior training efficiency compared to Search-R1, achieving peak performance in 80 training steps versus Search-R1's requirement of 120 steps—representing a 33\% efficiency enhancement. These findings validate that MCTS-guided exploration generates more informative learning signals, enabling more effective policy optimization for repository-level reasoning tasks.
  
  \end{tcolorbox}
  \label{fig:answer_rq1}
  \end{figure}

\subsection{RQ2: Benefits of Multi-turn Tool Reasoning for LLMs}

Given that our toolset is specifically designed for repository-level QA tasks, we evaluate whether LLMs can leverage multi-turn tool usage to enhance agent QA performance. Large-scale models such as Gemini and Claude possess strong tool utilization capabilities, and if our designed tools are effective, these models should demonstrate measurable performance improvements when employing the IRCoT method.

Tables~\ref{tab:closed_source_results} and~\ref{tab:open_source_comparison} present the evaluation results for both closed-source and open-source LLMs. For closed-source models, both Claude-3.5-Sonnet and Gemini-2.5-Pro demonstrate enhanced reasoning capabilities when employing multi-turn, tool-augmented IRCoT. Specifically, Gemini-2.5-Pro achieves a completeness score of 0.8729 with IRCoT, representing a 5.8\% improvement over naive generation. Claude-3.5-Sonnet shows even more substantial gains, with IRCoT achieving a score of 0.7800, corresponding to a 10.5\% improvement over naive generation. While GPT-4o's QA completeness with IRCoT does not surpass RAG performance, showing a -3.0\% decline compared to naive generation. 

Conversely, open-source models show mixed results with IRCoT: while Qwen3-14B achieves a modest 2.6\% improvement over naive generation, it still underperforms compared to RAG (11.1\% vs 2.6\%). Both Qwen3-32B (-6.7\%) and Qwen3-8B (-2.8\%) show performance declines compared to naive generation. These findings confirm that our toolset effectively enables large closed-source LLMs to explore code repositories, but smaller open-source models struggle with multi-turn tool reasoning without additional training.
\begin{table}[h]
  \centering
  \begin{minipage}[t]{0.5\textwidth}
  \centering
  \caption{Performance comparison of closed-source LLMs on repository QA tasks using different methods. Percentages indicate improvement over Naive Generation baseline.}
  \label{tab:closed_source_results}
  \vspace{-3mm}
  \resizebox{\textwidth}{!}{
  \begin{tabular}{c c c c}
  \toprule
  \textbf{Model} & \textbf{Method} & \textbf{$Comp_L$} & \textbf{vs. Naive Gen.} \\
  \midrule
  Claude-3.5-Sonnet & Naive Gen. & 0.7059 & - \\
                    & RAG & 0.7529 & +6.7\% \\
                    & IRCoT & \textbf{0.7800} & \textbf{+10.5\%} \\
  \midrule
  GPT-4o & Naive Gen. & 0.7482 & - \\
          & RAG & 0.7553 & +0.9\% \\
          & IRCoT & 0.7259 & -3.0\% \\
  \midrule
  Gemini-2.5 Pro & Naive Gen. & 0.8247 & - \\
                  & RAG & 0.8047 & -2.4\% \\
                  & IRCoT & \textbf{0.8729} & \textbf{+5.8\%} \\
  \bottomrule
  \end{tabular}
  }
  \end{minipage}
  \hfill
  \begin{minipage}[t]{0.48\textwidth}
  \centering
  \caption{Performance comparison of open-source LLMs on repository QA tasks using different methods. Percentages indicate improvement over Naive Generation baseline.}
  \label{tab:open_source_comparison}
  \vspace{-3mm}
  \resizebox{\textwidth}{!}{
  \begin{tabular}{c c c c}
  \toprule
  \textbf{Model} & \textbf{Method} & \textbf{$Comp_L$} & \textbf{vs. Naive Gen.} \\
  \midrule
  \qwenthirtytwob & Naive Gen. & 0.6541 & - \\
                  & RAG & 0.6647 & +1.6\% \\
                  & IRCoT & 0.6106 & -6.7\% \\
  \midrule
  \qwenthreeforteenb & Naive Gen. & 0.5388 & - \\
                     & RAG & 0.5988 & +11.1\% \\
                     & IRCoT & 0.5529 & +2.6\% \\
  \midrule
  \qwenthreeeightb & Naive Gen. & 0.5435 & - \\
                   & RAG & 0.5929 & +9.1\% \\
                   & IRCoT & 0.5282 & -2.8\% \\
  \bottomrule
  \end{tabular}
  }
  \end{minipage}
\end{table}

Conversely, Table~\ref{tab:open_source_comparison} reveals that open-source models, due to their smaller parameter scales, exhibit limited capacity for understanding complex tools and adapting reasoning based on tool outputs. Consequently, the code snippets retrieved through tool usage are often insufficient for answering queries, and multi-turn, tool-augmented reasoning fails to outperform RAG approaches. Some smaller models even underperform compared to direct generation, likely because ineffective tool usage introduces irrelevant information that degrades accuracy. This observation indicates that multi-turn tool-based reasoning requires models with sufficient parameter capacity to be effective without additional training.

\begin{figure}[h]
\centering
\vspace{-0.5em}
\begin{tcolorbox}[
  colback=gray!5,
  colframe=gray!50!black,
  title={\textbf{Answer to RQ2:}},
  fonttitle=\bfseries\color{white},
  coltitle=gray!50!black,
  rounded corners,
  boxrule=1pt,
  width=\textwidth
]
\small
Tool effectiveness strongly correlates with model scale. Large closed-source models (Gemini-2.5 Pro, Claude-3.5-Sonnet) benefit significantly from IRCoT, achieving consistent performance improvements. Conversely, smaller open-source models struggle with multi-turn tool interactions due to computational overhead exceeding benefits. This establishes a clear capability threshold: effective tool-based reasoning requires sufficient model capacity and reasoning sophistication.

\end{tcolorbox}
\vspace{-0.5em}
\label{fig:answer_rq1}
\end{figure}

\subsection{RQ3: MCTS Rollout Maintains Higher Entropy and Trajectory Diversity}

To address RQ3, we investigate the underlying mechanisms that enable MCTS-based exploration to achieve superior learning performance compared to general reinforcement learning strategies. Our analysis focuses on two key aspects: the evolution of cross-entropy during training and the diversity of sampled trajectories, which provide insights into the fundamental advantages of our approach.

We first introduce the concept of entropy in reinforcement learning training~\cite{cui2025entropy}. Entropy in RL reflects the LLM willingness to explore: higher entropy indicates a greater propensity to explore diverse action sequences and discover new reasoning paths, while lower entropy suggests the LLM becomes overly deterministic, repeatedly producing the same output. In reinforcement learning, it is desirable for the LLM to explore multiple paths so that rewards can distinguish between trajectories of varying quality, guiding policy updates toward higher-quality strategies. Once exploration collapses and trajectories converge to the same pattern, no comparative signal between trajectory qualities remains, leading to stagnation in policy improvement—a phenomenon often referred to as entropy collapse.

\begin{figure}[h]
  \centering
  \begin{subfigure}[b]{0.48\textwidth}
    \centering
    \includegraphics[width=\textwidth]{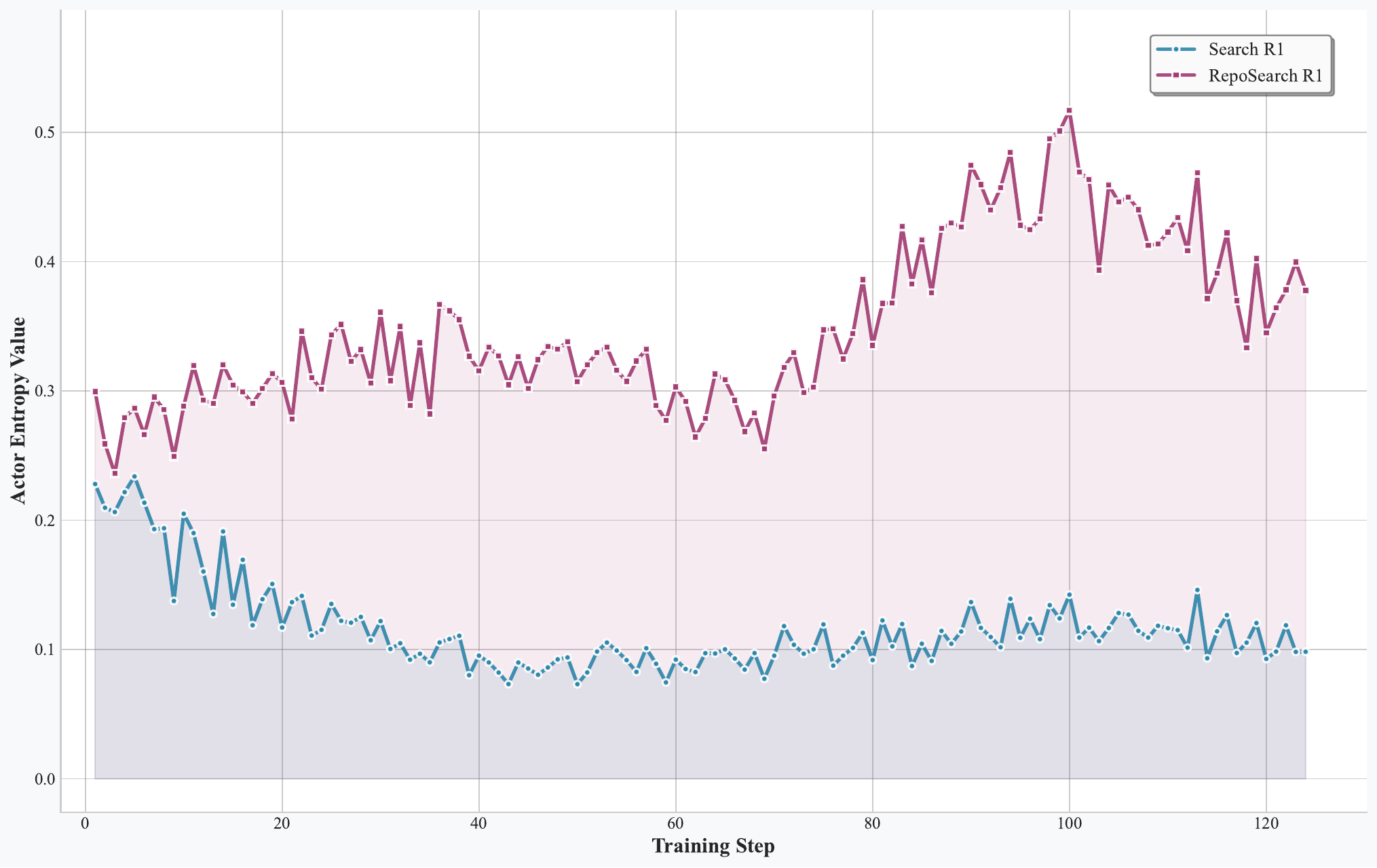}
    \caption{Actor Entropy Comparison}
    \label{fig:entropy}
  \end{subfigure}
  \hfill
  \begin{subfigure}[b]{0.48\textwidth}
    \centering
    \includegraphics[width=\textwidth]{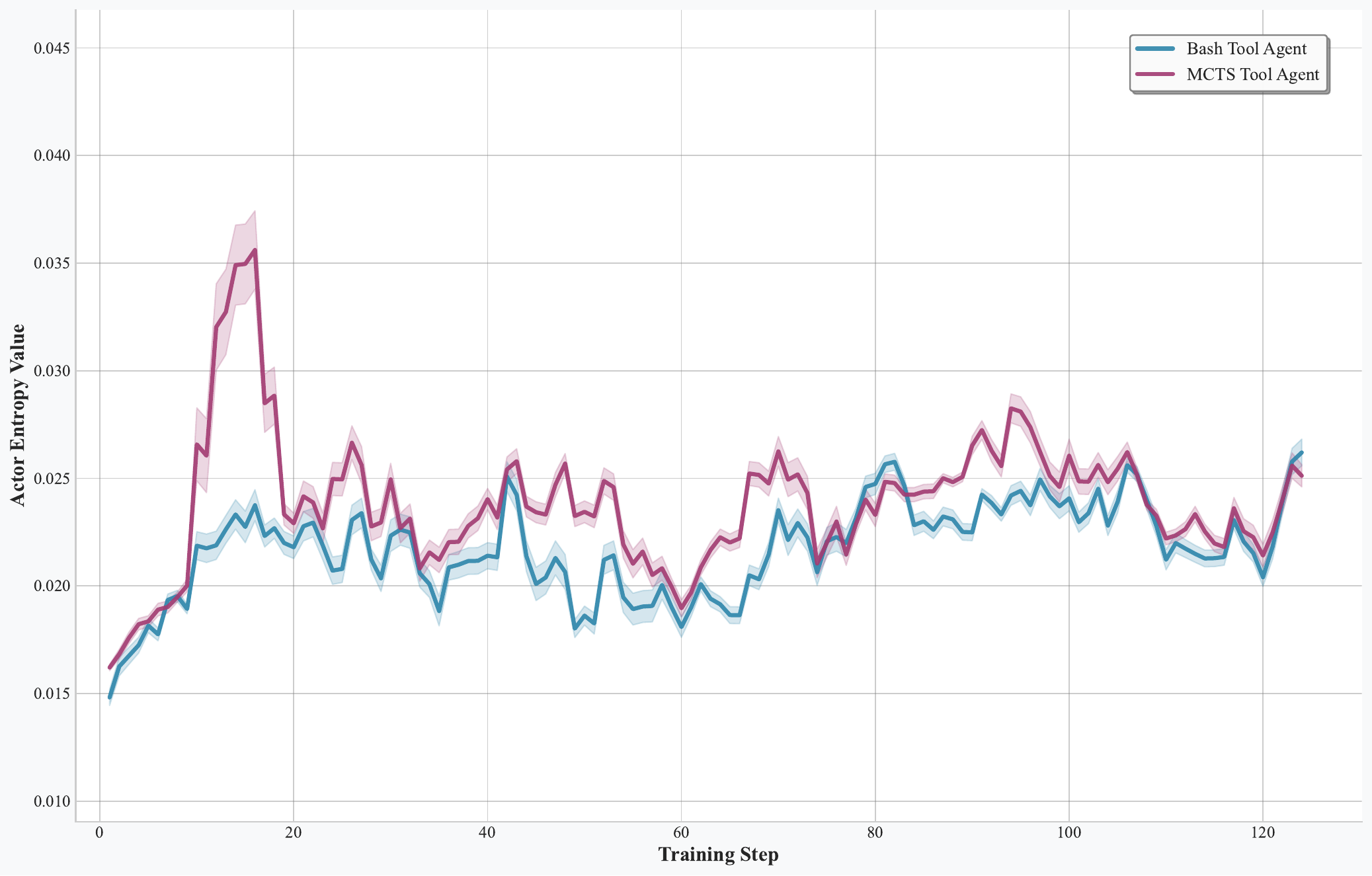}
    \caption{Trajectory Variance Comparison}
    \label{fig:traj_var}
  \end{subfigure}
  \caption{Comparison between Search-R1 and RepoSearch-R1: Entropy and Trajectory Variance Analysis. RepoSearch-R1 maintains consistently higher entropy levels with an exploration peak during mid-training, while generating higher-variance trajectories compared to Search-R1, demonstrating superior exploration diversity and preventing entropy collapse.}
  \label{fig:comparison}
\end{figure}

Several approaches have been proposed to mitigate entropy collapse, including DAPO~\cite{yu2025dapo} and Adaptive Entropy Control~\cite{he2025skywork}. Figure~\ref{fig:entropy} demonstrates the entropy evolution patterns for RepoSearch-R1 and Search-R1 throughout the training process. Notably, RepoSearch-R1 (purple curve) maintains consistently higher entropy levels and exhibits a distinctive exploration peak between training steps 40-60. Conversely, Search-R1's entropy rapidly deteriorates, reflecting increasingly deterministic behavior and diminished exploration capacity.
RepoSearch-R1's superior performance stems from its MCTS-based sampling mechanism, which constructs a Monte Carlo Thought Tree where UCT scores systematically guide reasoning node selection during exploration. This UCT-guided search transcends conventional sequential reasoning limitations, enabling systematic exploration of alternative solution pathways. Additionally, our self-critique-based child node generation actively promotes diverse action sequence discovery. The synergistic integration of these components sustains robust exploratory capabilities throughout the training process.

Figure~\ref{fig:traj_var} provides quantitative validation of this diversity advantage through trajectory variance analysis. RepoSearch-R1 consistently generates higher-variance trajectories (0.020-0.025 range) compared to Search-R1 (0.015-0.020 range). This elevated variance indicates that MCTS enables more diverse reasoning pathways and exploration strategies, substantially enriching training data quality. The sustained high trajectory variance throughout training validates that our exploration-decay UCT mechanism effectively balances exploration-exploitation trade-offs while preventing premature convergence to suboptimal policies.

\begin{figure}[h]
  \centering
  \begin{tcolorbox}[
    colback=gray!5,
    colframe=gray!50!black,
    title={\textbf{Answer to RQ3:}},
    fonttitle=\bfseries\color{white},
    coltitle=gray!50!black,
    rounded corners,
    boxrule=1pt,
    width=\textwidth
  ]
  \small
  MCTS-based self-training generates superior training data for repository search tasks through sustained exploration diversity. The consistently elevated entropy and trajectory variance demonstrate our method's capacity to produce diverse, high-quality trajectories, effectively addressing data scarcity challenges in repository-level reinforcement learning without requiring external supervision or model distillation.
  
  \end{tcolorbox}
  \label{fig:answer_rq1}
  \end{figure}

\section{RELATED WORK}
\label{sec:related_work}

\subsection{Agents for repository-level software engineering tasks}

The emergence of repository-level benchmarks, particularly SWE-bench~\cite{jimenez2023swe}, has established more realistic software engineering evaluation scenarios that better reflect real-world development challenges. Agent-based approaches, including SWE-agent~\cite{yang2024swe}, OpenHands~\cite{wang2024openhands}, Moatless Tools~\cite{antoniades2024swe}, and Marscode Agent~\cite{liu2024marscode}, provide sophisticated toolsets that enable autonomous repository exploration through iterative reasoning and feedback-driven refinement. However, these approaches typically rely on pre-trained models without specialized training for repository-level reasoning, limiting their effectiveness in complex scenarios.

\subsection{Tool-integrated Reasoning}  
Tool-integrated reasoning represents a fundamental paradigm where agents leverage external tools to enhance their problem-solving capabilities beyond pure language generation. Early works such as TORA~\cite{gou2023tora}, Star~\cite{zelikman2022star}, and AgentRefine~\cite{fu2025agentrefine} demonstrated the effectiveness of training models to reason with mathematical and computational tools. More recent advances, including ReTool~\cite{feng2025retool}, ToolRL~\cite{qian2025toolrl}, and search-r1~\cite{jin2025search}, have applied reinforcement learning to improve tool usage in multi-hop reasoning tasks. However, most existing approaches focus on general-purpose tools rather than domain-specific repository navigation and code analysis tools required for software engineering tasks.

\subsection{Training Software Agents}

Specialized training for software engineering agents bridges the gap between general LLMs and domain requirements. Works like SWE-Smith~\cite{yang2025swe}, SWE-fixer~\cite{xie2025swe}, Lingma-SWE-GPT~\cite{ma2024lingma}, and SWE-gym~\cite{pan2024training} acquire trajectory data through model distillation, introducing external dependencies and compliance concerns. SWE-RL~\cite{wei2025swe} employs self-enhancement without distillation but lacks sophisticated exploration strategies. Our work addresses these limitations through MCTS-based exploration for repository-level reasoning.

\section{DISCUSSION AND THREATS TO VALIDITY}
\label{sec:discussion}

\subsection{Efficiency of Agentic RL Framework}

We implemented RepoSearch-R1 using the veRL framework's Agent Loop architecture, which requires complete MCTS sampling across an entire batch before proceeding to policy probability calculations. This synchronous design creates training inefficiencies when question-answering tasks require varying completion steps, as batch processing becomes bottlenecked by slower samples—a prevalent long-tail problem in Agent RL. While our 500-sample training scale remained computationally manageable, large-scale deployment would face significant efficiency challenges. Recent asynchronous agent training frameworks offer promising solutions to this limitation. For example, ROLL~\cite{wang2025reinforcement} implements a rollout scheduler that immediately initiates feedback computation upon individual sample completion, eliminating dependencies on slower samples and mitigating long-tail effects.

\subsection{Effectiveness of Cold-Start Learning}

Our cold-start reinforcement learning framework specifically targets data compliance challenges in corporate environments, where external model distillation often violates data governance policies and introduces regulatory risks. The self-training MCTS mechanism successfully demonstrates that high-quality trajectory data can be generated entirely from internal resources, eliminating external dependencies while maintaining competitive performance. We acknowledge that in environments without compliance constraints, leveraging larger models for distillation with out-of-distribution (OOD) data could potentially yield superior behavioral patterns and more effective learning outcomes.

\subsection{Generalizability Limitations}
Our evaluation focuses exclusively on repository-level question-answering tasks within the CoReQA dataset, lacking validation across broader agent applications such as code generation, debugging, or multi-step software engineering workflows. This narrow task scope raises legitimate concerns regarding the generalizability of our approach to diverse agent applications beyond repository reasoning. Additionally, time constraints limited our exploration of alternative MCTS configurations beyond fundamental hyperparameter tuning. Future research should investigate comprehensive parameter optimization, including diverse exploration strategies, rollout policies, and tree expansion mechanisms, which could potentially enhance RepoSearch-R1's performance ceiling and unlock additional capabilities.

\subsection{Evaluation Bias and Reward Hacking}
Our evaluation methodology relies predominantly on LLM-judger for assessing answer quality and reasoning effectiveness, introducing inherent evaluation bias risks. Models may learn to generate responses that align with evaluator preferences rather than achieving genuine accuracy, creating a reward hacking scenario where outputs satisfy the judge without reflecting true correctness. This limitation becomes particularly problematic in reinforcement learning contexts where biased reward signals directly influence policy optimization trajectories. Future work should incorporate human review processes or ensemble evaluation methodologies that aggregate assessments across multiple models, thereby mitigating single-evaluator bias and ensuring more robust performance validation.

\section{CONCLUSION}
\label{sec:conclusion}

This paper presents RepoSearch-R1, a novel cold-start reinforcement learning framework that integrates Monte Carlo Tree Search (MCTS) with Group Relative Policy Optimization (GRPO) to enhance LLMs' repository-level reasoning capabilities. Our approach eliminates dependence on external model distillation through self-training with key innovations including exploration-decay UCT mechanism, self-critique-based child node generation, and dual reward structure. Experimental validation demonstrates substantial improvements: 19.4\% over IRCoT, 16.0\% over naive generation, and 6.4\% over RAG methods, with 33\% training efficiency gains.

This work demonstrates that sophisticated reasoning capabilities can emerge through self-supervised reinforcement learning without distilled data, opening new possibilities for training autonomous agents in complex environments where traditional supervised learning faces data scarcity.

\section{Data Availability}

We release our training source code of RepoSearch-R1 to encourage further exploration in this direction, while the CoReQA dataset could be found in~\cite{chen2025coreqa}. The artifact that supports the
results discussed in this paper is available at https://github.com/LingmaTongyi/RepoSearch-R1 ~\cite{repo}.

\newpage
\bibliographystyle{ACM-Reference-Format}
\bibliography{cite}

\end{document}